\documentstyle[aas2pp4]{article}

\lefthead{Nakano and Makino}
\righthead{Cusp around Black Holes in Luminous Ellipticals}

\begin{document}

\title{On the Cusp around Central Black Holes in Luminous Elliptical Galaxies}

\author{Taro Nakano}
\affil{Department of General Systems Studies, College of Arts and
Sciences, University of Tokyo, \\
3-8-1 Komaba, Meguro-ku, Tokyo 153-8902, Japan}
\authoremail{nakano@chianti.c.u-tokyo.ac.jp}

\and

\author{Junichiro Makino}
\affil{Department of Astronomy, School of Science, University of Tokyo\\
7-3-1 Hongo, Bunkyo-ku, Tokyo 113-0033, Japan}
\authoremail{makino@astron.s.u-tokyo.ac.jp}

\begin{abstract}
In this letter, we show that a massive black hole (MBH) which falls
into the center of a galaxy in dynamical timescale leaves a weak cusp
($\rho \propto r^{-1/2}$) around it, which is in good agreement with
the recent observations of luminous ellipticals by {\it Hubble Space
Telescope}. Such event is a natural outcome of merging of two galaxies
which have central MBHs. This is the only known mechanism to form weak
cusps in luminous ellipticals. Therefore, the existence of the weak
cusps indicates the central BHs of luminous ellipticals have fallen to
the center from outside, most likely during a major merger event.
\end{abstract}


\section{Introduction}
Recent observations of elliptical galaxies by {\it Hubble Space
Telescope} ({\it HST}) (\cite{lau95}; \cite{byu96}; \cite{geb96};
\cite{fab97}; \cite{kor96}) have brought us some challenging problems
as well as new knowledges on the structure of the central regions of
elliptical galaxies. First, these observations showed that no
elliptical has ``core'', in which the surface brightness or the
luminosity density profiles would be flat. The central regions
formerly regarded as ``cores'' proved themselves to be power-law cusps in all
observed galaxies. Second, the distribution of the slopes of these
cusps seem to be bimodal, one is the group of ``core'' galaxies with weak
cusps ($\rho \propto r^{-\alpha}, 0.4 \lesssim \alpha \lesssim 0.8$)
and the other is the group of  ``power-law'' galaxies with steep cusps
($\rho \propto r^{-\alpha}, \alpha \sim 2$). In addition, the slopes
of the cusps have correlation with the brightness of the galaxies, so
that bright galaxies tend to have weak cusps. Carollo et al. (1997)
claimed that the distribution of the slope is rather continuous in
their sample. However, their sample is limited in the range of
absolute magnitude, so it is not clear whether their result is real or 
due to selection.

The models of cuspy stellar systems previously studied are classified
into two categories --- models with and without MBH. However, neither
can explain the origin of the weak cusps in luminous elliptical
galaxies. It was shown that the cusp around BH would have the profile
$\rho \propto r^{-7/4}$ when the evolution is driven by the thermal
relaxation (\cite{bah76}; \cite{sha78}; \cite{coh78}; \cite{mar79};
\cite{mar80}) and $\rho \propto r^{-3/2}$ (\cite{you80}; \cite{mer98})
or rather steeper (\cite{qui95}) when the central BH grows
adiabatically. Hierarchical clustering in CDM cosmogony (\cite{nav96};
\cite{fuk97}) or dissipationless collapse (\cite{hoz99}) might form
relatively shallow cusps, but they are still significantly steeper
than the observed weak cusps.

Makino \& Ebisuzaki (1996, hereafter ME96) showed that a weak cusp
($\rho \propto r^{-\alpha}, \alpha \lesssim 1$) is formed through the
merging of two galaxies which have central BHs. They also found that
the ratio between the size of the weak cusp region and the half-mass
radius of the merger remnant, $r_{\rm c}/r_{\rm h}$, is proportional
to the ratio between the BH mass and the galaxy mass $M_{\rm
BH}/M_{\rm g}$. This result is in good agreement with observations
(\cite{geb96}). However, they did not discuss why the cusp was
formed. Quinlan \& Hernquist (1997) and Makino (1997) showed that the
BH binary in galactic center ejects many stars as it hardens and this
process can explain the weak cusps in large ellipticals, but they made
no quantitative prediction about cusp slopes.

\begin{figure}[htbp]
\plotone{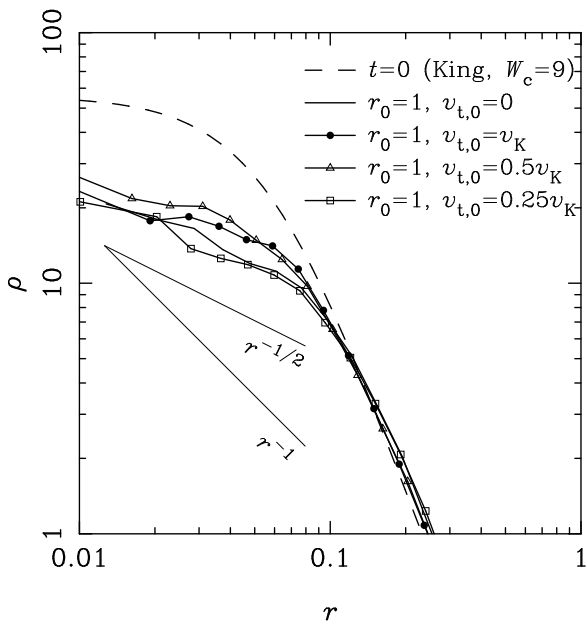}
\caption{Density profiles obtained by $N$-body simulations of BH
``fall-in'' (\protect\cite{nak99}). In these runs, the BH mass $M_{\rm
BH}=1/24 M_{\rm gal}$. BH is placed at distance of $r_{0}=1$ from the
center with tangential velocity $v_{\rm t,0}$. Note that $v_{\rm K}$
denotes the initial Kepler velocity. In all runs, initial radial
velocity of BH is zero. \label{fig:1}}
\end{figure}

In our previous work (\cite{nak99}, hereafter NM99), we investigated the
dynamical reaction of a galaxy to a BH which falls to the center, in
order to clarify the formation mechanism of the cusp. We found that when
the massive BH falls to the galaxy center, the stars are heated up
by the BH and the weak cusp ($\rho \propto r^{-1/2}$) is formed
(Figure \ref{fig:1}). This result is independent of the initial orbital
angular momentum and the mass of BH. Thus, we can conclude that when a 
BH (or BHs) falls from outside to the center of a galaxy, a central
weak cusp is always formed. However, it was not at all clear why the
cusp is formed.

In this letter, we present the theoretical explanation of the
formation mechanism of the weak cusp.

\section{Why the Cusp Tends to $r^{-1/2}$ ?}
\begin{figure}[htbp]
\plotone{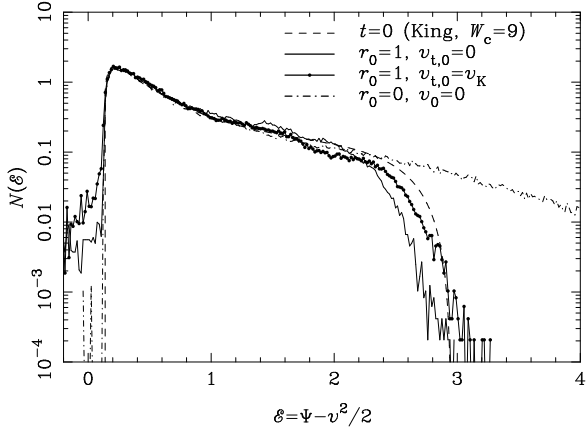}
\caption{The profiles of the binding energy distribution $N({\cal E})$
of the galaxy after the BH settled in the center, obtained by our
$N$-body calculation (\protect\cite{nak99}), where ${\cal E}$ is the
binding energy, ${\cal E}=\Psi - v^{2}/2$, and $\Psi$ is the potential
energy defined to be minus the conventional gravitational
potential. For all runs, the initial galaxy model is an isotropic King
model with non-dimensional central potential $W_{\rm c}=9$. The mass
of the BH, $M_{\rm BH}$, is $1/24$ of the mass of the galaxy. \label{fig:2}}
\end{figure}

In the results of our $N$-body simulation, we found an important
feature of the energy distribution function $N({\cal E})$ profiles,
shown in Figure \ref{fig:2}. The system of units we use is the standard unit
(\protect\cite{heg86}), in which the total mass of a galaxy $M_{\rm
gal} = 1$, the gravitational constant $G = 1$ and the total energy of
the galaxy $E_{\rm tot} = -1/4$. In this unit, the virial radius of
the galaxy is scaled to unity and the half-mass crossing time is
2$\protect\sqrt{2}$.

In Fig. \ref{fig:2}, $N({\cal E})$ is practically unchanged from
that of the initial King model for the runs in which the BH is
initially placed off-center. Thus, there is no star with ${\cal E} >
3.3$. This is in sharp contrast with $N({\cal E})$ for the run in
which the BH is initially placed on-center (dash-dot-dash line). In
this case, $N({\cal E})$ has a long tail to ${\cal E}
\rightarrow \infty$. By taking such depletion of tightly bound stars
into account, we can explain the existence of the weak cusp as
follows.

The density $\rho(r)$ is derived from the distribution function
$f({\cal E})$ as
\begin{equation}
\rho(r)=4\pi \int_{0}^{\Psi(r)} f({\cal E}) \sqrt{2\left[%
\Psi(r)-{\cal E} \right]} d{\cal E},
\label{eq:rho}
\end{equation}
where $\Psi(r)$ is the depth of the gravitational potential at
distance $r$ from the center and ${\cal E}$ is the specific binding
energy. Here we assume the galaxy is spherically symmetric and the
velocity distribution is isotropic. When the galaxy has the central BH,
$\Psi(r)$ near the BH can be approximated as $\Psi(r) \sim G M_{\rm
BH}/r$, and $\Psi(r)$ diverges when $r$ goes to zero. As shown in
Fig.\ref{fig:2}, $N({\cal E})$ of the galaxy after BH fell down to
the center vanishes at some finite value of the binding energy. We
denote this limit of binding energy as ${\cal E}_{\rm 0}$. Thus
$f({\cal E})$ also vanishes at ${\cal E}_{0}$, since $f({\cal
E})=N({\cal E})/A({\cal E})$ where $A({\cal E})$ is the area of
hypersurface in phase space with energy ${\cal E}$
(\cite{bin87}). Thus, if $\Psi(r) > {\cal E}_{\rm 0}$,
eq. (\ref{eq:rho}) can be rewritten as
\begin{equation}
\rho(r)=4\pi \int_{0}^{{\cal E}_{\rm 0}} f({\cal E}) \sqrt{2\left[%
\Psi(r)-{\cal E} \right]} d{\cal E}.
\label{eq:rho2}
\end{equation}

In the region $\Psi(r) > {\cal E}_{\rm 0}$, Eq. (\ref{eq:rho2}) can be
expanded as
\begin{eqnarray}
\rho(r) &=& 4\pi \int_{0}^{\cal E_{\rm 0}} f({\cal E}) \sqrt{2\left[%
\Psi(r)-{\cal E} \right]} d{\cal E} \nonumber \\
 &=& 4\sqrt{2}\pi \sqrt{\Psi(r)} \nonumber \\
 & & \times \int_{0}^{\cal E_{\rm 0}} f({\cal E}) \left[%
1-\frac{1}{2}\frac{{\cal E}}{\Psi(r)}+O\left({\left[\frac{{\cal
E}}{\Psi(r)}\right]}^{2} \right) \right] d{\cal E} \nonumber \\
& \propto & \sqrt{\Psi(r)} \sim \sqrt{\frac{G M_{\rm BH}}{r}}.
\end{eqnarray}
Therefore, in the central region where $\Psi(r) \gg {\cal E}$,
$\rho(r)$ is proportional to $r^{-1/2}$.

Note that our theory is related to, but not the same as, the theory
for $r^{-1/2}$ cusp by Zel'dovich \& Novikov (1971) and Peebles
(1972). They showed that if a massive object is placed in a stellar
system of uniform density (such as very large core), there will be
small cusp with $\rho \propto \sqrt{1+(GM_{\rm BH}m/rE_{\infty})}$, where
$E_{\infty}$ is the energy of a field star. In their theory, this cusp is
due to gravitational focusing of stars of uniform background. In our
theory, the cusp is also due to stars which are not bound to BH, but
we showed that uniform background is not necessary. Our explanation
needs only one assumption: $f({\cal E})$ vanishes at a certain finite
energy. This assumption is quite natural for the remnant of the
merging of galaxies with MBHs, since MBHs would heat up the stars
through the back reaction of the dynamical friction from the stars to
the falling MBHs.

\section{Self-consistent Model of Weak Cusp}
\begin{figure}[htbp]
\plotone{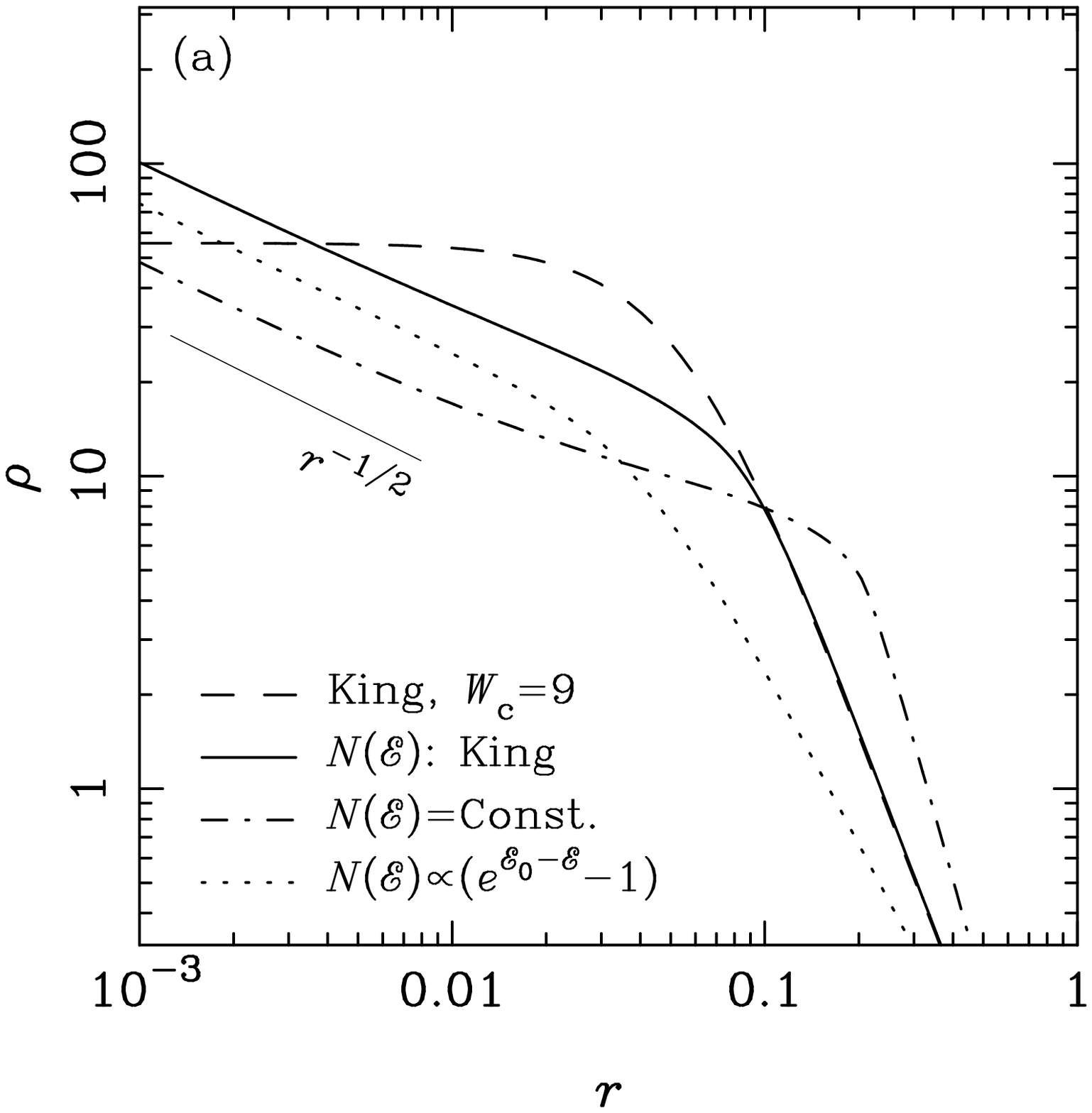}\\
\plotone{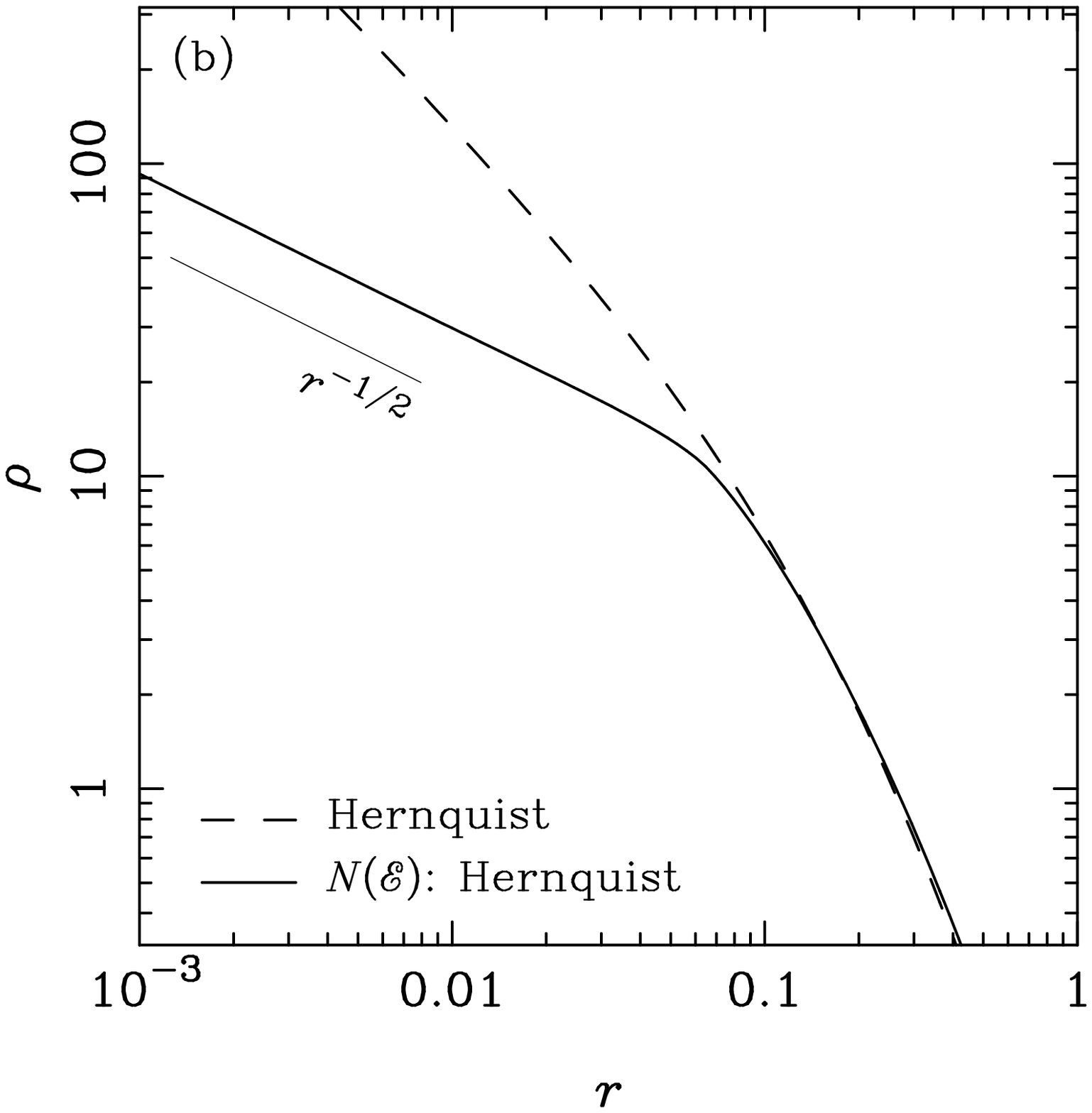}
\caption{(a) Self-consistent solutions of the density profile for various
$N({\cal E})$, obtained by the iterative method
(\protect\cite{bin82}). (b) Same results as (a) but using the
Hernquist model as initial galaxy model. \label{fig:3}}
\end{figure}

Figure \ref{fig:3}a shows the self-consistent solutions of the density
profiles of the galaxy with the central BH, for three different forms
of $N({\cal E})$; a King model with $W_{\rm c}=9$ (solid), a constant
$N({\cal E})$ (dot-dashed) and exponential with cutoff $N({\cal E})
\propto (e^{{\cal E}_{\rm 0}-{\cal E}}-1)$ (dotted), where ${\cal
E}_{\rm 0}$ is cutoff energy. The mass of BH is $1/24 M_{\rm gal}$. We
used the iterative method introduced by Binney (1982) to obtain these
self-consistent solutions. Figure \ref{fig:3}b is the same result as
Fig.\ref{fig:3}a but using the Hernquist model as initial guess of
iteration. In all cases, the steep outer slope switches to the weak
cusp with $\rho \propto r^{-1/2}$ at around $r = 0.1$. We can clearly
see that the difference in the functional form of $N({\cal E})$ does
not affect the slope of the central cusp. This result is in good
agreement with our analytical explanation described above.

\begin{figure}[htbp]
\plotone{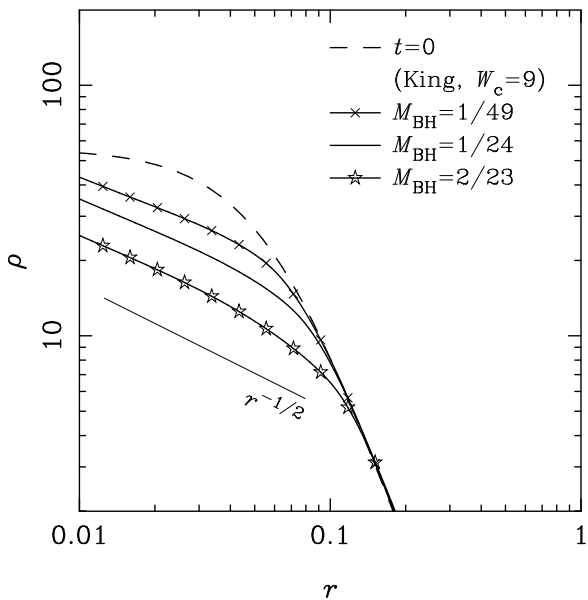}
\caption{Self-consistent density profiles with different BH
masses. The initial model is the King model with $W_{\rm
c}=9$. \label{fig:4}}
\end{figure}

Figure \ref{fig:4} shows the solutions of the density profiles with
different BH masses. The size
of the weak cusp region (or the so-called ``break radius'') is larger
for larger BH mass, but the slope of the weak cusp remains unchanged.

The relation between BH mass and size of weak cusp region (hereafter
simply term it ``core'') can be understood as follows. Suppose that the
inner region of initial power-law density profile $\rho = K_{1} r^{p}$
is transformed by sinking BH to a shallow cusp $\rho = K_{2} r^{q}$
with radius $r_{\rm c}$, where $-3 < p < q < 0$. Strictly speaking, a
King model has a small flat core, but we can neglect its contribution
in the following discussion. Roughly speaking, the 
total mass of the region affected by the BH must be about the same as
that of BH. Thus we have
\begin{equation}
M_{\rm BH} + \int_{0}^{r_{\rm c}} K_{2}r^{q} \cdot 4\pi r^{2}dr \simeq \int_{0}^{r_{\rm c}} K_{1}r^{p} \cdot 4\pi r^{2}dr.
\label{eq:potential}
\end{equation}
Using $K_{2}r_{\rm c}^{q}=K_{1}r_{\rm c}^{p}$, we obtain
\begin{equation}
r_{\rm c} \propto {M_{\rm BH}}^{1/(p+3)}.
\end{equation}
If the initial profile is isothermal ($p = -2$), the core radius would
be proportional to BH mass.

Such correlation between BH mass and core size is consistent with the
observations of ellipticals, which indicate that there are linear
correlations between the core radius and the effective radius
(\cite{fab97}) and between the central BH mass and total mass
(\cite{mag98}).

\section{Summary}
In this paper, we showed that the density cusp around MBH has the
slope of $-1/2$ if the MBH has ``fallen'' to the center in dynamical
timescale. Thus, we now understood why such weak cusps were formed in
numerical simulations of merging with central BHs (\cite{mak96}) or
simulation of sinking BHs (\cite{nak99}).

\begin{figure}[htbp]
\plotone{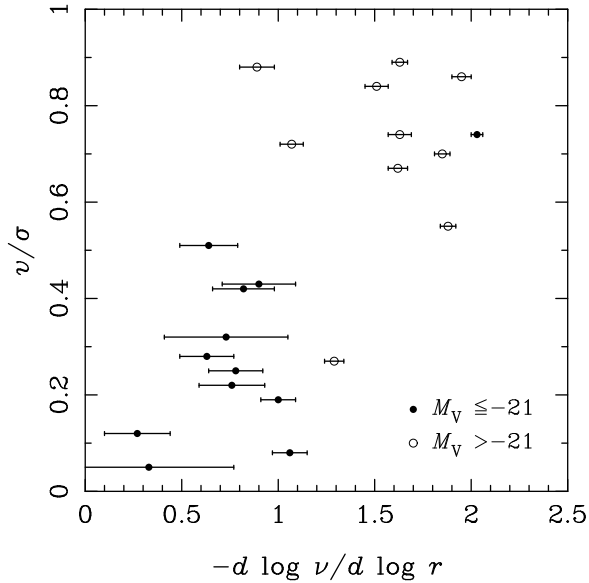}
\caption{Normalized rotation velocity of galaxies $v/\sigma$ versus
cusp slope of luminosity density profile at $r = 0.1^{\prime\prime}$
obtained by {\it HST} observations (\protect\cite{geb96};
\protect\cite{fab97}). Filled circles denote the galaxies brighter
than $M_{\rm V} = -21$ and open circles denote those fainter than
$M_{\rm V} = -21$. \label{fig:5}}
\end{figure}

This is the only known process
to form a weak cusp in the center of galaxies. Therefore, our result
very strongly suggests that luminous elliptical galaxies with weak
cusps have experienced such ``fall-in'' of MBHs. As suggested by ME96,
mergings of galaxies with MBHs already in the center is the most natural
scenario for such an event. In other words, luminous ellipticals are
most likely merger remnants. Our result also gives us an important
suggestion for the origin of central MBHs, which are believed to exist
in many galaxies (\cite{ric98}). The central MBHs in luminous
elliptical galaxies with weak cusps were not formed there by some process such as gas accretion but were imported dynamically from the progenitor galaxies
through the merging of them. If the central BH was formed in the
timescale longer than the dynamical timescale, the central density
cusp would have the slope of $-3/2$ or steeper
(\cite{bah76}; \cite{you80}). Therefore, the observed shallow cusp and the
existence of central MBHs are consistent only if ellipticals with
shallow cusps are merger remnants. The clear dichotomy of weak cusps
and steep cusps, and the correlation between the slope of the cusp and 
the normalized rotation velocity $v/\sigma$ (Figure \ref{fig:5})
suggests that not only the merger of gas-poor ellipticals but also the
``major merger'' (\cite{bar92}) of spirals (\cite{bar97}) resulted mostly
in weak cusps. Note that this clear separation is also visible in the
sample of Carollo et al (1997, Figure 8). This connection between
central slope and kinematics implies that in the
merging of two spirals, BHs would have become massive before two
galaxies finally merge, through gas-fueling induced by the tidal
torque (\cite{sel99}), and moreover, the rest of the gas would be
ejected or formed into stars more rapidly than the BHs sink to the
center of merger remnant.

\acknowledgments

We thank Yoko Funato for useful advice on the numerical method and for 
valuable discussions on our results. We are grateful to Toshiyuki
Fukushige for his comments on the calculation method and for
stimulating discussions. We thank Scott Tremaine for his important
comments on the previous works on this field. We also thank Daiichiro
Sugimoto and all the people who developed the special-purpose computer
GRAPE-4. The present work is supported in part by the Research for the
Future Program of Japan Society for the Promotion of Science,
JSPS-RFTP 97P01102.

\end{document}